\documentclass[prl,twocolumn,amsmath,amssymb,14pt,superscriptaddress]{revtex4-1}

\usepackage{fullpage}
\usepackage{psfrag}
\usepackage{graphicx}
\usepackage{xcolor}
\usepackage{subfigure}

\usepackage[latin1]{inputenc}
\usepackage{graphicx}
\usepackage{amsmath}
\usepackage{wasysym}
\usepackage{amssymb}
\usepackage{relsize}

\usepackage{soul}

\newcommand{\be}{\begin{equation}}
\newcommand{\ee}{\end{equation}}

\newcommand{\half}{\textstyle{\frac{1}{2}} \displaystyle}

\newcommand{\mean}[1]{\left \langle #1 \right \rangle}
\newcommand{\Fig}[1]{Fig.~\ref{#1}}
\newcommand{\Eq}[1]{Eq.~\ref{#1}}
\newcommand{\Tab}[1]{Table~\ref{#1}}

\newcommand{\Fs}{F_{\rm{s}}} 
\newcommand{\Fd}{F_{\rm{d}}} 
\newcommand{\Fc}{F_{\rm{c}}} 
\newcommand{\Fa}{F_{\rm{a}}}

\newcommand{\eps}{\epsilon}
\newcommand{\DF}{\Delta F}

\newcommand{\km}{\kappa_{\rm{m}}}
\newcommand{\kt}{\kappa_{\rm{t}}}
\newcommand{\ke}{\kappa_{\rm{eff}}}
\newcommand{\vz}{v_{0}}

\newcommand{\Cu}{C^{\rm{un}}}
\newcommand{\Ct}{C^{\rm{to}}}
\newcommand{\Nu}{N^{\rm{un}}}
\newcommand{\Nt}{N^{\rm{to}}}

\newcommand{\is}{\rm{s}^{-1}}
\newcommand{\s}{\rm{s}}
\newcommand{\pN}{\rm{pN}}
\newcommand{\nano}{\rm{nm}}

\begin{document}

\title{Force-dependent unbinding rate of molecular
  motors\\ from stationary optical trap data} \author{Florian Berger}
\email{fberger@rockefeller.edu}
\affiliation{Laboratory of Sensory Neuroscience, The Rockefeller
  University, New York, 10065 NY, USA} \author{Stefan Klumpp}
\affiliation{Institute for Nonlinear Dynamics, Georg-August University
  G{\"o}ttingen, 37077 G{\"o}ttingen, Germany} \author{Reinhard
  Lipowsky} \affiliation{Theory \& Bio-Systems, Max Planck Institute
  of Colloids and Interfaces, 14424 Potsdam, Germany} \date{\today}

\begin{abstract} {Molecular motors walk along filaments until they
    detach stochastically with a force-dependent unbinding rate. Here,
    we show that this unbinding rate can be obtained from the analysis
    of experimental data of molecular motors moving in stationary
    optical traps. Two complementary methods are presented, based on
    the analysis of the distribution for the unbinding forces and of
    the motor's force traces. In the first method, analytically
    derived force distributions for slip bonds, slip-ideal bonds, and
    catch bonds are used to fit the cumulative distributions of the
    unbinding forces. The second method is based on the statistical
    analysis of the observed force traces. We validate both methods
    with stochastic simulations and apply them to experimental data
    for kinesin-1.}
\end{abstract}

\pacs  {87.16.Nn, 87.15.Fh, 87.16.A-}

\maketitle

\paragraph{Introduction.}
In mammals, at least 80 genes code for different cytoskeletal motors
that transduce chemical free energy into mechanical work
\cite{howard2005, schliwa2003}. These molecular motors perform
nanometer steps along filaments from which they unbind stochastically
after a finite run length \cite{lipowsky2005}. Both their stepping
dynamics and their unbinding behavior are strongly affected by
external forces. In cells, these forces arise, e.g., from viscous
drag, from the elastic coupling to other force-producing molecules, or
from their cargo load \cite{hunt1994,rogers2009, coppin1997}. Thus,
stepping and unbinding are characterized by a force-velocity relation
and a force-dependent unbinding rate, respectively.

Our understanding of how different molecular motors respond to
external forces is primarily based on single-motor experiments with
optical traps \cite{veigel2011}. Whereas the force-velocity relations
have been studied for a variety of motors \cite{coppin1997,
  carter2005, schnitzer2000, clemen2005, gennerich2007}, the
force-dependent unbinding rate has been elucidated only for the
kinesin-1 motor \cite{andreasson2015, coppin1997, thorn2000}. The
motor-filament bond of kinesin behaves as a slip bond, i.e., the
unbinding probability increases with increasing load force.  In
contrast, the dynein motor may exhibit catch bond behavior, i.e., may
bind to the filament more strongly under force \cite{andreasson2015,
  rai2013}.  Furthermore, single dynein heads behave as slip-ideal
bonds, i.e., the unbinding rate first increases with force and then
becomes essentially force-independent \cite{nicholas2015}. Developing
reliable methods to determine the force-dependent unbinding behavior
of molecular motors, is important to advance our understanding of
their functions.
 
\begin{figure}[]
  \centering
\includegraphics[scale=1]{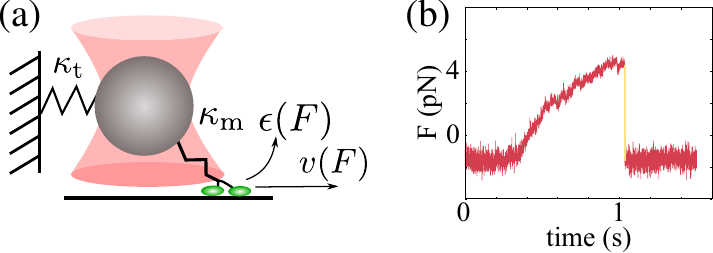}
\caption{\label{fig:1}(color online). {\bf Molecular motor in a stationary optical
    trap} (a) As the motor pulls the bead out of the trap center with
  velocity $v(F)$, the force $F$ on the motor increases, slows the
  motor down until it unbinds from the filament, and the bead falls
  back to the trap center. The stochastic unbinding is governed by the
  force-dependent unbinding rate $\eps(F)$. (b) The shape of the force
  trace depends on the motor dynamics and on the stiffnesses $\kt$ and
  $\km$ of trap and motor. The unbinding rate $\eps(F)$ is an
  independent motor parameter that reflects the molecular interactions
  between motor and filament.}
\end{figure}

In a standard setup, a single molecular motor pulls a bead against the
resisting force of a stationary optical trap \cite{coppin1997}. While
the motor moves away from the center of the trap, the force on the
bead increases until the motor unbinds from the filament and the bead
snaps back to the trap center, see \Fig{fig:1}. The force at which the
motor unbinds defines the unbinding force and a distribution of these
forces can be constructed from many such events.

In the present letter, we derive analytical expressions for the
unbinding force distributions.  Comparing these results to
experimental data allows us to identify the underlying filament-motor
bond behavior.  Furthermore, we estimate the force-dependent unbinding
rate with a complementary approach based on the statistical analysis
of force traces \cite{coppin1997,thorn2000}. The latter method uses
the information of the whole trace and does not require prior
knowledge of the motor's elastic properties or of its force-velocity
relation.  We explicitly show how both methods are connected and
discuss their limitations.  After validating both methods with
stochastic simulations, we apply them to experimental data to
determine the force-dependent unbinding behavior of kinesin-1.

\paragraph{Distribution-based method.}
The first method is based on the distribution $p(F)$ of the unbinding
forces as measured experimentally or in simulations. To derive analytical
expressions for the distribution, we extend a method previously used to 
analyze force-spectroscopic data of single molecules \cite{dudko2006,
  dudko2008}. This method transforms the distribution of
unbinding forces into the force-dependent unbinding rate
\begin{equation}
\label{eq:epsbasic}
  \epsilon(F) = \frac{\dot{F}(F) \, p(F)}{1-\int_0^F p(F')dF'},
\end{equation}
in which $\dot{F}(F)$ is the force-dependent loading rate, i.e., the
rate at which the force changes.  From (\ref{eq:epsbasic}), we obtain
the probability distribution function (pdf) 
\be
  \label{eq:pFm}
  p(F)=\frac{\epsilon(F)}{\dot{F}(F)}\exp\left[-\int_0^F\frac{\epsilon(F')}{\dot{F}(F')}
  dF'\right] 
\ee
for the unbinding force. 
The latter equation implies that the distribution $p(F)$ is determined by
the ratio of  unbinding rate $\eps(F)$ to  loading rate
$\dot{F}(F)$. Therefore, from the pdfs for the unbinding force, we can
only estimate the ratio $\eps/\dot{F}$, but not the unbinding rate
alone without knowing the loading rate. 

However, we can estimate
$\dot{F}$ from a theoretical description of the motor. In the simplest
case, the motor has a linear force-velocity relation
$v(F) \equiv \vz (1-F/\Fs)$, in which $\Fs$ is the stall force and
$\vz$ the load-free velocity \cite{klumpp2015}. The force-extension
relation of the motor molecule is assumed to be linear with spring constant
$\km$.  This spring is connected in series with the spring-like
potential of the optical trap described by the spring constant
$\kt$. Using this motor description, we obtain the
force-dependent loading rate $\dot{F}(F) = \ke \vz(1-F/\Fs)$ with 
the effective stiffness $\ke \equiv \km \kt/(\km + \kt)$. First,  we assume that
the motor exhibits slip-bond behavior with unbinding rate 
$\eps(F) \equiv \eps_0 \exp(F/\Fd)$, the detachment force $\Fd$ and the load-free
unbinding rate $\eps_0$. For such an unbinding rate, the relation 
(\ref{eq:pFm}) leads to the unbinding force distribution 
 \begin{equation}
\label{eq:pfexp}
p(F) = \frac{\Fs \epsilon_0 e^{F/\Fd}}{\ke v_0 (\Fs-F)} 
\exp \left[-\frac{\Fs \epsilon_0  }{\ke v_0} \Phi(F)                   \right]
\ee
with
\be
\Phi(F) \equiv    e^{\Fs/\Fd}
\left(I\left(\frac{\Fs-F}{\Fd}\right)-I\left(\frac{\Fs}{\Fd}\right)\right)
\ee
and  the exponential integral function $I(x)\equiv \int_x^{\infty}
t^{-1}\exp[-t] dt$. In a similar way, we calculate exact expressions
for the distribution of unbinding forces for slip-ideal and catch
bonds \cite{berger2020}. These expressions can be used to fit empirical cumulative distributions constructed from data, thereby
estimating the unbinding rate $\eps(F)$. To validate our distribution-based approach, 
we use stochastic simulations to generate data sets of unbinding
forces for different types of motor-filament bonds \cite{berger2020}. 

\begin{figure}[t!]
\includegraphics[scale=0.95]{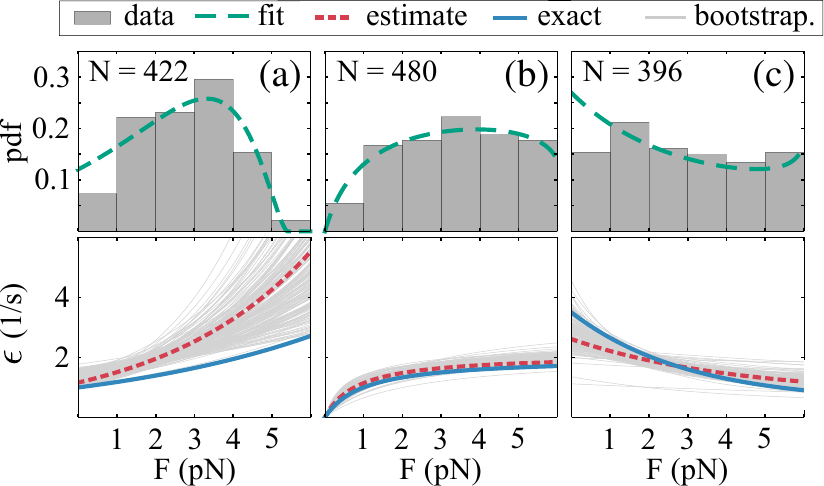}
\caption{\label{fig:2}(color online). {\bf Distribution-based estimate of the
    unbinding rate (simulation study)} From a stochastic simulation of
  a molecular motor in an optical trap, we obtain a set of unbinding
  forces from which we construct an empirical cumulative distribution
  function \cite{berger2020}. We fit these cumulative distributions by
  the cumulative distributions as obtained from the analytically
  derived unbinding-force distributions (or pdfs) $p(F)$. Optimizing
  these fits, we obtain the distributions displayed as green lines,
  which are in good agreement with the gray normalized histograms for
  (a) slip bonds, (b) slip-ideal bonds, and (c) catch bonds. The
  corresponding unbinding rate $\eps(F)$ are the red lines in the
  lower panels. For comparison, we also display the unbinding rates
  used to generate the data (blue lines).  The gray lines in the lower
  panels illustrate the variability of the unbinding rate from
  bootstrapping \cite{berger2020}.}
\end{figure}

To account
for the experimental noise in the trajectories, we add an appropriate
level of noise to the simulated data. In this way,  we generate
three different data sets for slip, slip-ideal, and catch bonds. We choose the
parameter values for the three different bonds such that
the resulting unbinding rates have a comparable numerical range, see
\Fig{fig:2}. We then use the analytically derived expressions for the unbinding-force
distributions to fit the empirical cumulative distributions of the simulated data to deduce the
parameters of the unbinding rates. The three different force-dependent
unbinding behaviors lead to distinct unbinding-force distributions,
see green lines in \Fig{fig:2}. The estimated parameters
are in good agreement with the parameters used to
generate the data \cite{berger2020}. However, this agreement reflects, to a
large extend,   our knowledge about the assumed functional forms
of the unbinding rates and of the dynamics of the motor. Next, we
describe a method that uses the whole ensemble of force traces to estimate the
unbinding rate without assuming any microscopic model.

\paragraph{Trace-based method.}
To estimate the force-dependent unbinding rate from experimental data,
without any assumptions about the motor-filament bonds and the 
force-velocity relation,  it is necessary to estimate both the loading
rate $\dot{F}$ and the distribution $p(F)$ of unbinding forces, see
\Eq{eq:epsbasic}. The loading rate $\dot{F}$ can be estimated from the slope of the force traces before unbinding and $p(F)$ 
can be obtained as a 
histogram of the unbinding forces \cite{dudko2008}. The histogram has $N$ bins with bin width
$\DF$. The height $h_i$ of  the $i$-th bin is determined from the counts $C_i$
per bin as $h_i = C_i/(\Nu\DF)$ with $\Nu \equiv \sum_i C_i$. An estimator for the force-dependent
unbinding rate  in (\ref{eq:epsbasic}) is then given by \cite{dudko2008} 
\begin{equation}
\label{est_unbind}
  \eps[(k-\half)\DF]= \frac{\dot{F}[(k-\frac{1}{2})\DF] h_k}{\DF(\frac{h_k}{2} +
  \sum_{i=k+1}^N h_i)}.
\end{equation}
This equation involves the force-dependent loading rate
$\dot{F}[(k-\frac{1}{2})\DF]$ which has to be determined from the
slopes of the force traces. We
rewrite (\ref{est_unbind})  in such a way that the unbinding rate can
be estimated directly from the data points of the force traces
\cite{berger2020}. We bin the data points of all force traces into $N$
force bins with bin width $\DF$ and label $k$. We determine the numbers $\Cu_k$ of
unbinding events and the number $\Ct_k$ of data points of all
force traces per bin. The force-dependent unbinding rate is then given
by
\begin{equation}
\label{eq:eps}
  \eps[(k-\half)\DF] = \frac{\Cu_k}{\delta t \, \Ct_k},
\end{equation}
in which $\delta t$ is the time step between the recorded points of
the trace. Thus $\delta t \, \Ct_k$ represents the total time of all
force traces spent in the $k$-th bin. To evaluate this equation
neither a microscopic model nor any kind of fitting procedure is
needed. Applying this approach to our simulated data, we determine the
underlying force-dependent unbinding rate solely from the
force traces, see \Fig{fig:3}. Even though the distributions of the
unbinding forces for the slip-ideal and the catch bond appear to be similar,  
see \Fig{fig:2}, the trace-based method distinguishes the
different unbinding behaviors remarkably well.

\begin{figure}[t!]
  \centering
\includegraphics[scale=1]{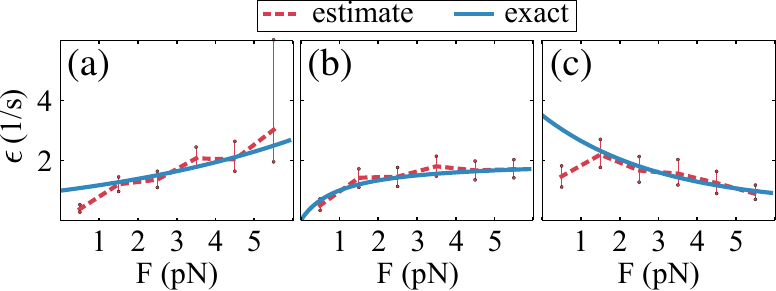}
\caption{\label{fig:3}(color online). {\bf Trace-based estimate of the
    unbinding rate (simulation study)} Without assuming a microscopic
  model, we only use the unbinding forces and the force traces from
  the stochastic simulations to estimate the unbinding rate for the
  three bond behaviors: slip bond (a), slip-ideal bond (b),
  and catch bond (c). The estimated unbinding rates (red lines)
  are in very good agreement with the exact unbinding rates used for
  the simulations (blue lines). The 95\% confidence intervals are
  obtained from bootstrapping \cite{berger2020}.}
\end{figure}

%{\color{red}*** revised up to here, Mo, April 9, 7:45 Uhr ***}

\paragraph{The unbinding rate of kinesin-1.}

We apply both methods to experimental data of kinesin-1 pulling a bead
out of a stationary optical trap. The data was obtained during control experiments carried out for a previous
study \cite{deberg2013}. The force-free velocity $v_0\simeq 484\, \nano /\s$ and the trap stiffness $\kt \simeq 0.03 \,
\pN/\nano$ \cite{deberg2013}. We assume a motor stiffness 
of $\km \simeq 0.3 \, \pN/\nano$ \cite{coppin1997}. However,
this assumption is not crucial because the effective
stiffness $\ke$ is dominated by the much smaller trap stiffness $\kt$. We
determine the unbinding rate from a fit of the empirical
cumulative distribution function constructed from 682 unbinding events, see
\Fig{fig:4}(a,b) and \cite{berger2020}. Despite our simplifying assumptions,  the fit is in good
agreement with the data, indicating that kinesin's
unbinding rate is consistent with a slip-bond behavior. We find the following optimal parameters with
confidence intervals given in brackets: $\eps_0 \simeq 0.97 \,[0.80;
1.35]\, \is$, $\Fd \simeq 2.25\,[2.03; 5.18] \, \pN$, and $\Fs \simeq
14.97 \, [6.26; 15]\, \pN$. Using the trace-based approach, we
determine the force-dependent unbinding rate of kinesin-1 from \Eq{eq:eps}
as shown in
\Fig{fig:4}(d). An exponential fit excluding the boundary points  
leads to a detachment force of $\Fd \simeq 7.4 \, \pN$ and a force-free
unbinding rate of $\eps_0 \simeq 1.1 \, \is$.

\begin{figure}[t!]
  \centering
\includegraphics[scale=0.85]{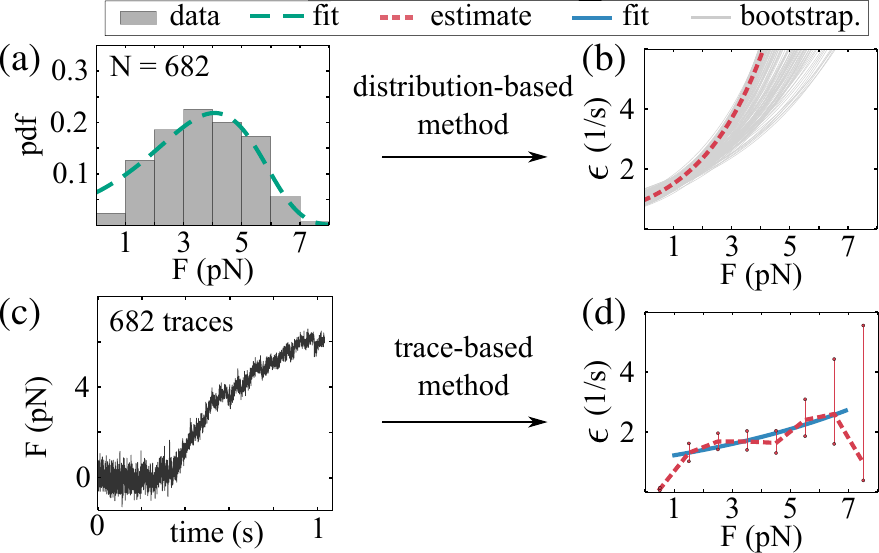}
\caption{\label{fig:4}(color online). {\bf Unbinding rate of kinesin-1 (experimental
    data)} (a,b) Distribution-based method: We use the analytical
  expression in \Eq{eq:pfexp} for a slip bond to fit the empirical
  cumulative distribution constructed from the experimental data
  \cite{berger2020}. The numerical values of the fitted parameters
  determine the unbinding-force pdf $p(F)$ (green line in panel a) and
  the unbinding rate $\eps(F)$ (red line in panel b). For comparison,
  the distribution of the experimental data is estimated by a gray
  histogram. (c,d) Trace-based method: We use all 682 force traces
  \textemdash one example shown in (c) \textemdash to obtain the
  unbinding rate $\eps(F)$ as the red line in (d) from \Eq{eq:eps}.
  Fitting this trace-based estimate with an exponential function (blue
  line), we obtain the force-free unbinding rate $\eps_0 \simeq 1.1 \,
  \is$ and the detachment force $\Fd \simeq 7.4 \, \pN$. In (b), the
  gray lines illustrate the variability of the unbinding rate from
  bootstrapping; in (d) the errors are given as 95\% confidence
  intervals \cite{berger2020}.{\color{red}}}
\end{figure}

\paragraph{Discussion.}
We have explicitly derived the distribution of unbinding forces of a
single molecular motor in a stationary optical trap and validated it
by stochastic simulations. The  trace-based method reliably infers the correct
unbinding behavior from the simulated data, see \Fig{fig:3}.

Furthermore, we have shown that a simple description of kinesin-1 is
consistent with the experimental distribution of unbinding
forces. However, the consistency implies a smaller detachment force
compared to the estimate obtained from the trace-based
approach. In conclusion, our trace-based analysis suggests a
force-free unbinding rate of $\epsilon_0 \simeq 1.1 \, \is $ and a
detachment force of $\Fd \simeq 7.5 \, \pN$. While the unbinding rate
is consistent with the value of $\epsilon_0 \simeq 1.0 \, \is $
commonly used for kinesin-1, the detachment force is 2.5 times larger
than the value of $\Fd \simeq 3 \, \pN$ used in most theoretical
studies \cite{klumpp2015}. However, recent experimental and
modeling studies indicate a higher value of about $ 6 - 7 \, \pN$ or
even a more complicated behavior \cite{andreasson2015, goekerarpa2014,
  sumi2017}.

Force-dependent unbinding has important consequences for
the function of the motors in  their cellular environment
\cite{chaudhary2018, blehm2013, berger2011}. Theoretical descriptions
based on single-molecule dynamics indicate that many   emerging
phenomena, such as cooperative transport and macroscopic force
production can only be explained with a suitable force-dependent
unbinding rate for the single motors \cite{berger2012,klumpp2005-a,
  mueller2008, berger2017}. 

Our framework provides a systematic way to study the force-dependent
unbinding rate of molecular motors and can be extended to describe
more complex optical trapping experiments. To determine the unbinding
rate for forces that exceed the stall force, the stage could be moved
relative to the trap which adds only an extra term to the loading
rate \cite{nicholas2015}. 

A first step towards understanding the function of motor proteins is
to determine biophysical quantities that are directly accessible to
experiments \cite{ruhnow2017}. 
While the probability distribution of unbinding forces
depends on the stiffness of the trap and also on the stiffness of the
linker that connects the motor to the bead, the force-dependent
unbinding rate is a characteristic property of the motor-filament
bond. Therefore, our computational approach provides a systematic framework for future studies to
distinguish different motor-filament bonds such as dynein's catch bond from kinesin's slip bond.

\paragraph{Acknowledgment.}
We thank Paul Selvin, Hannah A. DeBerg and Benjamin H. Blehm for
providing the experimental data and stimulating discussions. The
experimental data was acquired under the support of the NIH. FB was
supported by a grant from the Alexander von Humboldt-Foundation.

\clearpage

\onecolumngrid
\begin{center}
  \textbf{\large Supplementary Materials: Force-dependent unbinding rate of
  molecular motors from stationary optical trap data}\\[.5cm]
  Florian Berger,$^{1}$ Stefan Klumpp,$^{2}$ and Reinhard
  Lipowsky$^3$\\[.1cm]
  {\itshape ${}^1$Laboratory of Sensory Neuroscience, The Rockefeller
  University, New York, 10065 NY, USA\\
  ${}^2$Institute for Nonlinear Dynamics, Georg-August University
  G{\"o}ttingen, 37077 G{\"o}ttingen, Germany\\
  ${}^3$Theory \& Bio-Systems, Max Planck Institute
  of Colloids and Interfaces, 14424 Potsdam, Germany\\}
(Dated: \today)\\[1cm]
\end{center}
%\twocolumngrid

\setcounter{equation}{0}
\setcounter{figure}{0}
\setcounter{table}{0}
\setcounter{page}{1}
\renewcommand{\theequation}{S\arabic{equation}}
\renewcommand{\thefigure}{S\arabic{figure}}
\renewcommand{\bibnumfmt}[1]{[S#1]}
\renewcommand{\citenumfont}[1]{S#1}

\section{Distribution of unbinding forces}
For our distribution-based analysis we derive analytic expressions
for the distribution of unbinding forces for three different
force-dependent bond behaviors.

\subsection{The slip bond}
We describe a slip bond with an unbinding rate that increases with the external force as
\begin{equation}
  \eps(F) \equiv \eps_0 \exp(F/\Fd).
\end{equation}
Here, we introduce the force-free unbinding rate $\eps_0$ and the
characteristic detachment force $\Fd$. To calculate the corresponding
distribution of unbinding forces, we proceed as explained in the main
text and find
\begin{equation}
  p(F)= \frac{\Fs \epsilon_0 \exp[F/\Fd]}{\ke v_0 (\Fs-F)}
  \exp\left[-\frac{\epsilon_0 \exp[\Fs/\Fd] \Fs}{\ke v_0} \left(I\left(\frac{\Fs-F}{\Fd}\right)-I\left(\frac{\Fs}{\Fd}\right)\right)\right],
\end{equation}
with the exponential integral function
\begin{equation}
  I(x)\equiv \int_x^{\infty} t^{-1}\exp[-t] dt.
\end{equation}
This equation implies that we cannot obtain $v_0$, $\ke$ and
$\epsilon_0$ independently from a fit, only the combination
\begin{equation}
\label{eq:Fc}
  \Fc \equiv \frac{v_0 \ke}{\eps_0} 
\end{equation}
that defines the characteristic force $\Fc$. We simplify the
distribution to
\begin{equation}
\label{eq:slipP}
  p(F|\Fc,\Fd,\Fs)= \frac{\Fs \exp[F/\Fd]}{\Fc(\Fs-F)}
  \exp\left[-(\Fs/\Fc) \exp[\Fs/\Fd] \left(I\left(\frac{\Fs-F}{\Fd}\right)-I\left(\frac{\Fs}{\Fd}\right)\right)\right],
\end{equation}
which depends now on the three parameters, $\Fc$, $\Fd$, and $\Fs$. 

\subsection{The slip-ideal bond}
An ideal bond is characterized by a constant unbinding rate that is
independent of the force \cite{dembo1994s, nicholas2015s}. As a
phenomenological description for a slip-ideal bond with an unbinding
rate that first increases with force and then becomes constant, we use
the rational function
\begin{equation}
  \label{eq:slipideal}
  \eps(F) \equiv \frac{\eps_0 F}{\Fa+F}.
\end{equation}
The corresponding probability distribution for the unbinding forces
follows as
\begin{equation}
\label{eq:idealP}
p(F|\Fa,\Fc,\Fs)=  \frac{F \Fs}{\Fc(\Fa+F)(\Fs-F)}\exp\left(\frac{\Fs}{\Fc(\Fa+\Fs)}\left(\Fa
    \ln\left(1+\frac{F}{\Fa}\right) + \Fs \ln\left(1-\frac{F}{\Fs}\right)\right)\right),
\end{equation}
in which $\Fc$ is the characteristic force given in \Eq{eq:Fc}.

\subsection{The catch bond}
A catch bond is characterized by an unbinding rate that decreases with
increasing force \cite{dembo1994s}. We describe such a bond with the
force-dependent unbinding rate
\begin{equation}
  \label{eq:catch}
  \eps(F) = \eps_0\exp(-F/\Fd) + a.
\end{equation}
The corresponding distribution of unbinding forces reads 
\begin{equation}
\begin{aligned}
  \label{eq:catchP}
  p(F|\Fa,\Fc,\Fd,\Fs) & =\\ & \frac{\Fs(\Fc/\Fa + \exp[-F/\Fd])}{\Fc(\Fs-F)}
  \exp\biggl[(\Fs/\Fc)\exp[-\Fs/\Fd] \left(I\left(-\frac{\Fs}{\Fd}\right)-I\left(\frac{F-\Fs}{\Fd}\right)\right)+
  \\ & (\Fs/\Fa) (\ln(1-F/\Fs)\biggr)\biggr],
\end{aligned}
\end{equation}
in which $\Fc$ is defined in \Eq{eq:Fc} and $\Fa\equiv v_0 \ke/a$.

\section{Simulations}
To generate data for the validation of our methods, we assume that the
motor steps forward with $\ell = 8 \, \nano$ steps with a force-dependent
stepping rate $\alpha(F)$. We relate the stepping rate
 \begin{equation}
  \alpha(F) = v(F)/\ell = v_0(1- F/\Fs)/\ell
\end{equation}
to a linear force velocity $v(F)$ described by the stall force $\Fs
\simeq 6 \, \pN$
and a typical force-free velocity $v_0 \simeq 1 \, \mu\rm{m/s}$. For studying the different unbinding
behaviors we use the corresponding unbinding rates $\eps(F)$
introduced above and listed in \Tab{tab:1}. The stepping and unbinding
of the motor are force-dependent and the force on the motor changes as
it pulls the bead out of the center of the optical trap. To determine
the force exerted on the motor, we assume a linear restoring force
of the optical trap, characterized by a typical trap stiffness of
$\kt\simeq 0.01 \, \pN/\nano$ and a
linear force-extension relation for the motor molecule with spring
constant $\km\simeq 0.3 \, \pN/\nano$ \cite{coppin1997s}. The force
\begin{equation}
  F = \kt x_{\rm{b}}
\end{equation}
on the bead is determined from the distance $x_{\rm{b}}$ of the bead to the
center of the trap. This distance changes with the position $x_{\rm{m}}$ of
the motor, while it is moving out of the center of the trap, as
\begin{equation}
  x_{\rm{b}}=\frac{\km}{\km+\kt}x_{\rm{m}}.
\end{equation} 

Under the assumption that after each step the system reaches
mechanical equilibrium instantaneously, the force on the bead equals
the force on the bond of the motor and the filament with the
corresponding loading rate
\begin{equation}
  \dot{F}=\kt\frac{\km}{\km+\kt}\dot{x}_{\rm{m}}.
\end{equation} 

We base our stochastic simulation on a Gillespie algorithm: at each
step we calculate the force acting on the motor, adjust the stepping
and unbinding rates accordingly and choose the next event with the
corresponding probability \cite{gillespie1977s}. If the motor steps
forward, we increase the position $x_{\rm{m}}$ by the step size
$\ell$. If the motor unbinds from the filament, we set the position to
$x_{\rm{m}} = 0$. In this way, we obtain trajectories of a single
molecular motor as it pulls the bead out of a stationary trap. We
convert the spatial trajectories corresponding to the bead position
as a function of time into forces traces by multiplying
the bead position with the stiffness of the trap.  To mimic the
experimental force traces as closely as possible, we add Gaussian
white noise with a standard deviation of $\sigma = 0.3 \, \pN$, as
estimated from the experimental data. An example of such a force trace
is shown in \Fig{fig:s1}(a).

\begin{figure}[t!]
  \centering
\includegraphics[scale=0.27]{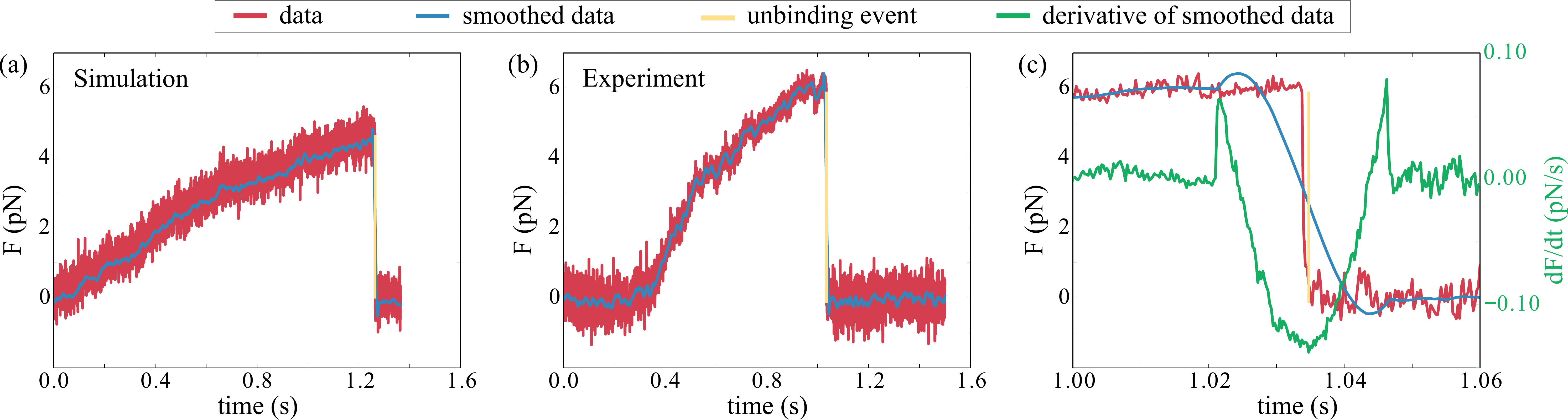}
\caption{\label{fig:s1} {\bf Analysis of force traces} The original
  data (red) from the simulations and experiments are smoothed with a
  Savitzky-Golay filter (blue) and one example for each are shown in
  (a) and (b) respectively. Our unbinding detection algorithm
  determines the magnitude of the unbinding force (yellow) and
  displays it close to the unbinding event to allow for visual
  inspection. To explain the basic idea of the algorithm, we magnified
  the unbinding event of (b) in (c) and show the derivative as the discrete
  difference (green) between points of the smoothed force
  trace. The location of the two maxima of the derivative are reliable
  estimates of time points before and after the unbinding
  event. Taking the difference of the averages of the traces before
  and after the unbinding event provides the numerical value for the
  unbinding force.}
\end{figure}

\section{Detection of unbinding forces from the  force traces}
The experiments and the simulations provide long force traces with
hundreds of binding and unbinding events. From these traces we
separate each force-generation event with its associated unbinding
event into a separate file, see \Fig{fig:s1}. We smooth the traces
with a Savitzky-Golay filter and take the average of the first 200
points to determine the baseline that we then subtract from the
trace. The base line subtraction is not necessary for the simulated
traces. From the smoothed trace the position of the unbinding event is
automatically detected in the following way; First, we estimate the
derivative of the trace as the finite difference between adjacent
points, see \Fig{fig:s1}(c). Then, the jump of the trace after the
unbinding event is identified at the time with the smallest
derivative. We estimate the numerical value for the unbinding force as
the difference of the trace before and after the jump. To obtain an
exact value, we need to estimate the time when the unbinding event
occurs and when the bead is equilibrated in the center of the
trap. Therefore, we start at the time at which the derivative is
negative and determine the two nearest time points when the derivative
is positive. One point corresponds to a time before the unbinding
event and the other point indicates that the bead is equilibrated at
the center of the trap. We average 20 data points before the first
time point and 20 data points after the equilibration. The difference
of these averages provides the numerical value of the unbinding
force. We visually inspect each trace and monitor the results of the
detection algorithm. In this way we obtain the set of all unbinding
forces.

\begin{figure}[t!]
  \centering
\includegraphics[scale=0.27]{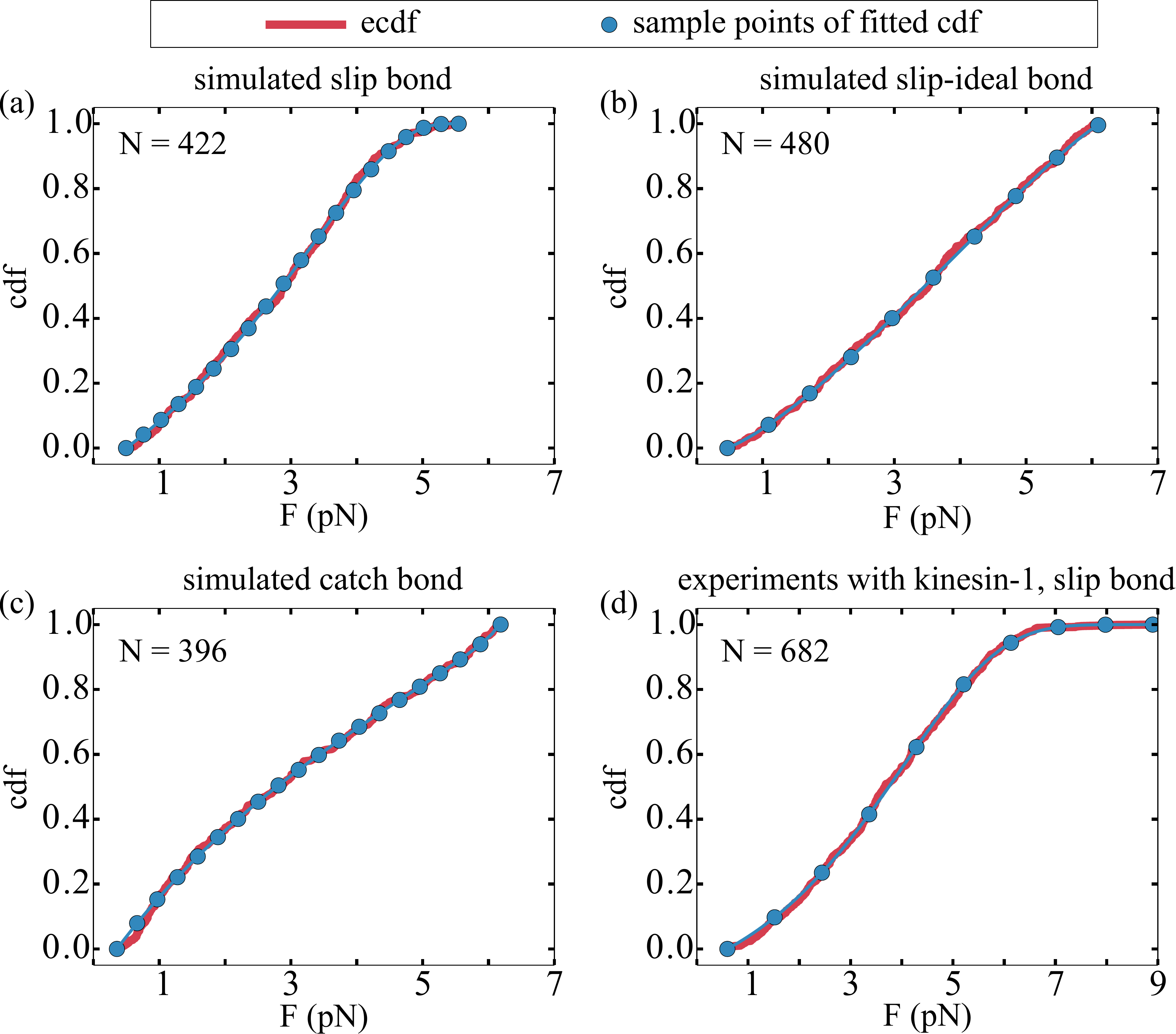}
\caption{\label{fig:s2} {\bf Fitted cumulative distributions:} The red
  lines represent the empirical cumulative distribution functions
  (ecdf) constructed from simulation data of (a) slip bonds, (b)
  slip-ideal bonds, (c) catch bonds, and (d) from experimental data
  for kinesin-1. As explained in the text, we fit each ecdf by the
  cumulative distribution function of \Eq{eq:cdfm} for the respective
  bond behavior. To enhance the minimization algorithm, we evaluate
  the integrals of the cdf at a finite number of sample points shown
  as the blue dots.}
\end{figure}

\section{Fitting of the distributions}
To determine the numerical values of the free parameters for a specific
unbinding behavior, we fit the analytic expressions of the
unbinding force distributions to either experimental or simulated
data. Because fitting a distribution directly to a histogram
constructed from the data depends on the arbitrary choice of the
number of bins, we instead fit the cumulative distribution functions
to empirical cumulative distribution functions (ecdf) constructed
from the data. For a data set of $n$ data points, the ecdf is a step
function that increases by $1/n$ at each of the $n$ data points. The
cumulative distribution function (cdf) is defined as
\begin{equation}
\label{eq:cdf}
  P(F) \equiv \int_{0}^F p(F')\,dF'.
\end{equation}
$P(F)$ is a probability and attains values between 0 and 1.
To account for the limited resolution of detecting unbinding events in
the experiments, we shift and renormalize the cdf with respect to the
smallest detected unbinding force $F_{\rm{min}}$ of each data set to  
\begin{equation}
\label{eq:cdfm}
  P_{\rm{m}}(F) \equiv \frac{P(F)-P(F_{\rm{min}})}{1-P(F_{\rm{min}})}.
\end{equation}
For our fitting procedure, we evaluate the integrals in the cdfs only
for a finite number of sample points $\{F_i\}$. In the case of the
slip-bond data with $N=422$ detected unbinding events, we find the
optimal numerical values of the parameters from the minimization
\begin{equation}
\min_{\Fc, \Fd, \Fs}{\sum_i\left({\rm ecdf}(F_i) - P_{\rm{m}}(F_i|\Fc,\Fd,\Fs)\right)^2},  
\end{equation}
in which $P_{\rm{m}}(F|\Fc,\Fd,\Fs)$ is given from combining \Eq{eq:slipP},
\Eq{eq:cdf}, and \Eq{eq:cdfm}. To enhance the
performance of the minimization routine in Mathematica, we restrict
the intervals for the parameters to $\Fc \in (0,+\infty)\,\pN$, $\Fd \in
(0,10)\,\pN$, and $\Fs \in (0,10)\,\pN$. We evaluate the integrals at 20
equidistant sample points $\{F_i\}$, see blue dots in \Fig{fig:s2}. From
this fitting procedure, we obtain $\Fc \simeq 8.32 \, \pN$, $\Fd
\simeq 3.85 \, \pN$, and $\Fs \simeq 5.41 \, \pN$. With $v_0 \simeq
1000\, \nano/s$, $\km \simeq 0.3 \, \pN/\nano$ and $\kt \simeq 0.01 \,
\pN/\nano$, we determine the force-free unbinding rate
\begin{equation}
 \eps_0 =\frac{v_0 \km \kt}{\Fc(\km+\kt)} \simeq 1.16 \, \is.  
\end{equation}
We obtain 95\% confidence intervals with a significance of $0.05$ from a bootstrapping procedure. We re-sample 200 data sets of the
original size $N=422$. For each set we determine the optimal fit
parameters and calculate the lower and upper confidence interval as
the 2.5 percentile and the 97.5 percentile of the distribution of each
fit parameter respectively. All parameter values are listed in \Tab{tab:1}.

We proceed in the same way for the slip-ideal bond for which we
identify $N=489$ unbinding events in the simulated data. During the
minimization, we restrict the parameters to $ \Fa \in (0.5,10)\,\pN$, $\Fc
\in (1,5)\,\pN$, and $\Fs \in (1,10)\,\pN$. We evaluate the integrals at 10
equidistant sample points $\{F_i\}$, see blue dots in \Fig{fig:s2}. The
numerical values for the optimal parameters are listed in \Tab{tab:1}
with confidence intervals calculated in the same way as for the slip
bond.

In the case of the catch bond, we determine the fit parameters as
explained for the slip bond above. Our simulated data set contains
$N=396$ unbinding events. To enhance the minimization, we restrict the
parameter to: $\Fa \in (0,20)\,\pN$, $\Fc \in (0,20)\,\pN$, $\Fd \in (0,20)\,\pN$,
and $\Fs \in (0,20)\,\pN$. We evaluate the integrals at 20 equidistant
sample points $\{F_i\}$, see blue dots in \Fig{fig:s2}. The optimal
numerical values for the free parameters are listed in \Tab{tab:1}
with the 95\% confidence intervals determined as for the two other
cases.

\begin{table}
  \centering
\begin{tabular}{l |r|c|c| c| }
bond behavior & parameter & value for the simulation & value from fit & CI \\
\hline
\hline
slip bond&
$\epsilon_0\, (\is)$  & 1 &1.16 & [0.91; 1.62] \\
 $\eps(F) = \epsilon_0 \exp(F/\Fd)$&$\Fd\, (\pN)$ & 6  & 3.85 & [2.51; 10.00] \\
&$\Fs\, (\pN)$ & 6   & 5.41 & [4.76;6.45]\\
\hline
\hline
slip-ideal bond&
$\epsilon_0 \, (\is)$ &2  &2.11 & [1.94; 3.06] \\
 $\eps(F) = \epsilon_0 F/(\Fa+F)$&$\Fa \, (\pN)$ &1  & 0.84 & [0.5; 5.78] \\
&$\Fs \, (\pN)$ &6   & 6.13 & [3; 6.28]\\
\hline
\hline
catch bond&
$\epsilon_0 \, (\is)$ &3  &1.86 & [0.86; 3.17] \\
 $\eps(F) = \epsilon_0 \exp(-F/\Fd) + a$&$a \, (\is)$ &0.5  & 0.75 &
 [0.48; 1.51] \\
&$\Fd \, (\pN)$ &3   & 4.18 & [0.63; 10.34]\\
&$\Fs \, (\pN)$ &6   & 6.19 & [2.28; 6.26]\\
\hline
\end{tabular}
\caption{\label{tab:1} {\bf Validation of distribution-based method:}
  Each bond behavior is described by the unbinding rate $\eps(F)$ as
  given in the first column, depending on three or four parameters as
  listed in the second column. We choose numerical values for these parameters to generate
  data from our stochastic simulation. By fitting the analytic
  expressions for the unbinding force distributions to the
  generated data, we estimate the numerical values. The 95\% confidence intervals are obtained from bootstrapping.}
\end{table}

\section{Trace-based method}
We estimate the force-dependent unbinding rate from the traces binned
into $N$ force bins with bin width $\DF$. We label the bins with $k$,
count the number $\Cu_k$ of unbinding events and the number $\Ct_k$ of
points of all traces in bin $k$. Intuitively, $\Ct_k$ is the total
number of possible unbinding events and $\Cu_k$ is the number of
actual unbinding events. Thus, the ratio
\begin{equation}
\label{eq:p}
\frac{\Cu_k}{\Ct_k}
\end{equation}
gives the probability of unbinding in the $k$-th bin at each sampling
point in time. We divide this expression by the time step $\delta t$ between the
samples to obtain the unbinding rate 
\begin{equation}
\label{eq:eps}
  \eps((k-0.5)\DF) = \frac{\Cu_k}{\delta t \Ct_k}.
\end{equation}
Note, the denominator $\delta t \Ct_k$ is equal to the total time of
all traces spent in the $k$-th bin. This expression has been
used previously to estimate the unbinding rate of molecular motors
\cite{coppin1997s, thorn2000s}. In the following, we rewrite this
estimator for the unbinding rate to obtain the estimator introduced by
Dudko et al. \cite{dudko2008s}.  The number of unbinding events per bin
is related to the density histogram $h_k$ as
\begin{equation}
\label{eq:pF}
  \Cu_k = \Nu \DF h_k,
\end{equation}
in which $\Nu$ is the total number of all unbinding events.
To approximate the total time that the traces spent in the $k$-th bin, we
first look at all traces that pass through the bin without
unbinding. The number of these traces is given by all
traces that unbind after bin $k$, i.e.,
\begin{equation}
  \sum_{i=k+1}^N \Cu_i.
\end{equation}
Multiplying this number by the mean time spent in the bin provides the
total time of all traces passing through that bin as
\begin{equation}
  \delta t \mean{\Nt_k}\sum_{i=k+1}^N \Cu_i,
\end{equation}
in which $\mean{\Nt_k}$ is the mean number of data points of a
trace in the bin. To account for traces that unbind at a
point in $\DF$, we assume that they unbind uniformly in the
interval $\DF$ and therefore the total time of these traces is
given by
\begin{equation}
  \frac{1}{2}\delta t \mean{\Nt_k} \Cu_k.
\end{equation}
Note that the factor $1/2$ accounts for the premature unbinding and
$\mean{\Nt_k}$ is the mean number of data points of all traces
that pass through the bin.

Taken together, the total time spent in bin $k$ reads
\begin{equation}
\label{eq:t}
  \delta t \Ct_k =   \delta t \mean{\Nt_k} \left (\frac{1}{2} \Cu_k +
  \sum_{i=k+1}^N \Cu_i \right ).
\end{equation}

To estimate the loading rate, we approximate the slope of the
trace at the center of the bin as constant per bin and obtain
\begin{equation}
  \dot{F}((k-1/2)\DF) = \frac{\DF}{\delta t \mean{\Nt_k}},
\end{equation}
from which we get
\begin{equation}
\label{eq:dN}
    \delta t \mean{\Nt_k}= \frac{\DF}{\dot{F}((k-1/2)\DF)}.
\end{equation}
Combining \Eq{eq:dN}, \Eq{eq:t}, \Eq{eq:pF} and \Eq{eq:eps}, we derive
the estimator for the unbinding rate as
\begin{equation}
\eps((k-0.5)\DF) = \frac{\dot{F}((k-1/2)\DF) h_k}{\DF(h_k/2 +
  \sum_{i=k+1}^N h_i)},  
\end{equation}
which is the estimator proposed by Dudko in \cite{dudko2008s}.

To obtain the unbinding rate with the trace-based method, we cut all
traces at their unbinding events. We then bin all points of all traces
according to their force and determine the number $\Ct_k$ of data
points in each bin $k$. The number $\Cu_k$ of unbinding events readily
follows from binning the set of unbinding forces. Together with the
inverse sampling rate $\delta t^{-1}$, we obtain an estimate for the
force-dependent unbinding rate from \Eq{eq:eps}.

\section{Kinesin-1 data}
To determine the force-dependent unbinding rate for kinesin-1, we cut
out $682$ single unbinding events together with the raising pulling
phase. We apply the trace-based method to this data set as explained
in the proceeding section. The results are discussed in the main text
of the manuscript. To apply the distribution-based method to the
experimental data, we determine the cumulative distribution for the slip
bond by combining \Eq{eq:slipP} and \Eq{eq:cdfm}. For the fitting
procedure we evaluate the integrals at 10 equidistant sample points,
see blue dots in \Fig{fig:s2}. We obtain the following optimal parameters
with their 95\% confidence intervals in brackets: $\Fc \simeq 15.62
\,[11.26; 18.9] \, \pN$, $\Fd \simeq 2.25 \,[2.03;5.18] \, \pN$ and
$\Fs \simeq 14.97 \,[6.26; 15]\, \pN$. To determine the force-free
unbinding rate $\eps_0$ from \Eq{eq:Fc}, we use the experimental
values $v_0 \simeq 484 \, \nano/\s$ and $\kt \simeq 0.03 \,
\pN/\nano$ \cite{deberg2013s}. The stiffness of the molecular motor is
not a crucial parameter, because the effective stiffness of the motor
and the trap is governed by the smaller trap stiffness. We assume a
numerical value of $\km \simeq 0.3 \, \pN/\nano$
\cite{coppin1997s}. With these parameter values, we obtain the
force-free unbinding rate $\eps_0 \simeq 0.97 \,[0.80; 1.35]\, \is$.

\section{Estimation of variability}
\subsubsection{Distribution-based method}
To illustrate the variability of the data, we re-sample 200 data sets
of the unbinding events and apply the fitting procedure. From each
calculation we obtain a force-dependent unbinding rate. In Fig. 4 of
the main text, we display the 95\% of the closest unbinding rates as
thin gray lines in the background.

\subsubsection{Trace-based method}
To determine the 95\% confidence intervals at a significance of
$0.05$ we use a bootstrapping approach. We re-sample 200
data sets of unbinding events with the corresponding traces of the
original size. For each data set we determine the unbinding rate
from \Eq{eq:eps}. Thus, we have for each sampling point of the force 200
different numerical values for the unbinding rate. We then calculate
the lower and upper confidence interval as the 2.5 percentile and the
97.5 percentile of the unbinding rates at each force step.


\begin{thebibliography}{30}%
\makeatletter
\providecommand \@ifxundefined [1]{%
 \@ifx{#1\undefined}
}%
\providecommand \@ifnum [1]{%
 \ifnum #1\expandafter \@firstoftwo
 \else \expandafter \@secondoftwo
 \fi
}%
\providecommand \@ifx [1]{%
 \ifx #1\expandafter \@firstoftwo
 \else \expandafter \@secondoftwo
 \fi
}%
\providecommand \natexlab [1]{#1}%
\providecommand \enquote  [1]{``#1''}%
\providecommand \bibnamefont  [1]{#1}%
\providecommand \bibfnamefont [1]{#1}%
\providecommand \citenamefont [1]{#1}%
\providecommand \href@noop [0]{\@secondoftwo}%
\providecommand \href [0]{\begingroup \@sanitize@url \@href}%
\providecommand \@href[1]{\@@startlink{#1}\@@href}%
\providecommand \@@href[1]{\endgroup#1\@@endlink}%
\providecommand \@sanitize@url [0]{\catcode `\\12\catcode `\$12\catcode
  `\&12\catcode `\#12\catcode `\^12\catcode `\_12\catcode `\%12\relax}%
\providecommand \@@startlink[1]{}%
\providecommand \@@endlink[0]{}%
\providecommand \url  [0]{\begingroup\@sanitize@url \@url }%
\providecommand \@url [1]{\endgroup\@href {#1}{\urlprefix }}%
\providecommand \urlprefix  [0]{URL }%
\providecommand \Eprint [0]{\href }%
\providecommand \doibase [0]{http://dx.doi.org/}%
\providecommand \selectlanguage [0]{\@gobble}%
\providecommand \bibinfo  [0]{\@secondoftwo}%
\providecommand \bibfield  [0]{\@secondoftwo}%
\providecommand \translation [1]{[#1]}%
\providecommand \BibitemOpen [0]{}%
\providecommand \bibitemStop [0]{}%
\providecommand \bibitemNoStop [0]{.\EOS\space}%
\providecommand \EOS [0]{\spacefactor3000\relax}%
\providecommand \BibitemShut  [1]{\csname bibitem#1\endcsname}%
\let\auto@bib@innerbib\@empty
%</preamble>
\bibitem [{\citenamefont {Howard}(2005)}]{howard2005}%
  \BibitemOpen
  \bibfield  {author} {\bibinfo {author} {\bibfnamefont {J.}~\bibnamefont
  {Howard}},\ }\href@noop {} {\emph {\bibinfo {title} {Mechanics of Motor
  Proteins and the Cytoskeleton}}},\ \bibinfo {edition} {2005th}\ ed.\
  (\bibinfo  {publisher} {Sinauer},\ \bibinfo {address} {Sunderland, Mass},\
  \bibinfo {year} {2005})\BibitemShut {NoStop}%
\bibitem [{\citenamefont {Schliwa}\ and\ \citenamefont
  {Woehlke}(2003)}]{schliwa2003}%
  \BibitemOpen
  \bibfield  {author} {\bibinfo {author} {\bibfnamefont {M.}~\bibnamefont
  {Schliwa}}\ and\ \bibinfo {author} {\bibfnamefont {G.}~\bibnamefont
  {Woehlke}},\ }\href {\doibase 10.1038/nature01601} {\bibfield  {journal}
  {\bibinfo  {journal} {Nature}\ }\textbf {\bibinfo {volume} {422}},\ \bibinfo
  {pages} {759} (\bibinfo {year} {2003})}\BibitemShut {NoStop}%
\bibitem [{\citenamefont {Lipowsky}\ and\ \citenamefont
  {Klumpp}(2005)}]{lipowsky2005}%
  \BibitemOpen
  \bibfield  {author} {\bibinfo {author} {\bibfnamefont {R.}~\bibnamefont
  {Lipowsky}}\ and\ \bibinfo {author} {\bibfnamefont {S.}~\bibnamefont
  {Klumpp}},\ }\href
  {http://econpapers.repec.org/article/eeephsmap/v_3a352_3ay_3a2005_3ai_3a1_3ap_3a53-112.htm}
  {\bibfield  {journal} {\bibinfo  {journal} {Physica A: Statistical Mechanics
  and its Applications}\ }\textbf {\bibinfo {volume} {352}},\ \bibinfo {pages}
  {53} (\bibinfo {year} {2005})}\BibitemShut {NoStop}%
\bibitem [{\citenamefont {Hunt}\ \emph {et~al.}(1994)\citenamefont {Hunt},
  \citenamefont {Gittes},\ and\ \citenamefont {Howard}}]{hunt1994}%
  \BibitemOpen
  \bibfield  {author} {\bibinfo {author} {\bibfnamefont {A.~J.}\ \bibnamefont
  {Hunt}}, \bibinfo {author} {\bibfnamefont {F.}~\bibnamefont {Gittes}}, \ and\
  \bibinfo {author} {\bibfnamefont {J.}~\bibnamefont {Howard}},\ }\href
  {\doibase 10.1016/S0006-3495(94)80537-5} {\bibfield  {journal} {\bibinfo
  {journal} {Biophysical Journal}\ }\textbf {\bibinfo {volume} {67}},\ \bibinfo
  {pages} {766} (\bibinfo {year} {1994})},\ \bibinfo {note} {{PMID:}
  7948690}\BibitemShut {NoStop}%
\bibitem [{\citenamefont {Rogers}\ \emph {et~al.}(2009)\citenamefont {Rogers},
  \citenamefont {Driver}, \citenamefont {Constantinou}, \citenamefont
  {Jamison},\ and\ \citenamefont {Diehl}}]{rogers2009}%
  \BibitemOpen
  \bibfield  {author} {\bibinfo {author} {\bibfnamefont {A.~R.}\ \bibnamefont
  {Rogers}}, \bibinfo {author} {\bibfnamefont {J.~W.}\ \bibnamefont {Driver}},
  \bibinfo {author} {\bibfnamefont {P.~E.}\ \bibnamefont {Constantinou}},
  \bibinfo {author} {\bibfnamefont {D.~K.}\ \bibnamefont {Jamison}}, \ and\
  \bibinfo {author} {\bibfnamefont {M.~R.}\ \bibnamefont {Diehl}},\ }\href
  {\doibase 10.1039/B900964G} {\bibfield  {journal} {\bibinfo  {journal}
  {Physical Chemistry Chemical Physics}\ }\textbf {\bibinfo {volume} {11}},\
  \bibinfo {pages} {4882} (\bibinfo {year} {2009})}\BibitemShut {NoStop}%
\bibitem [{\citenamefont {Coppin}\ \emph {et~al.}(1997)\citenamefont {Coppin},
  \citenamefont {Pierce}, \citenamefont {Hsu},\ and\ \citenamefont
  {Vale}}]{coppin1997}%
  \BibitemOpen
  \bibfield  {author} {\bibinfo {author} {\bibfnamefont {C.~M.}\ \bibnamefont
  {Coppin}}, \bibinfo {author} {\bibfnamefont {D.~W.}\ \bibnamefont {Pierce}},
  \bibinfo {author} {\bibfnamefont {L.}~\bibnamefont {Hsu}}, \ and\ \bibinfo
  {author} {\bibfnamefont {R.~D.}\ \bibnamefont {Vale}},\ }\href
  {http://www.ncbi.nlm.nih.gov/pmc/articles/PMC23000/} {\bibfield  {journal}
  {\bibinfo  {journal} {Proceedings of the National Academy of Sciences of the
  United States of America}\ }\textbf {\bibinfo {volume} {94}},\ \bibinfo
  {pages} {8539} (\bibinfo {year} {1997})},\ \bibinfo {note} {{PMID:} 9238012
  {PMCID:} {PMC23000}}\BibitemShut {NoStop}%
\bibitem [{\citenamefont {Veigel}\ and\ \citenamefont
  {Schmidt}(2011)}]{veigel2011}%
  \BibitemOpen
  \bibfield  {author} {\bibinfo {author} {\bibfnamefont {C.}~\bibnamefont
  {Veigel}}\ and\ \bibinfo {author} {\bibfnamefont {C.~F.}\ \bibnamefont
  {Schmidt}},\ }\href {\doibase 10.1038/nrm3062} {\bibfield  {journal}
  {\bibinfo  {journal} {Nature Reviews Molecular Cell Biology}\ }\textbf
  {\bibinfo {volume} {12}},\ \bibinfo {pages} {163} (\bibinfo {year}
  {2011})}\BibitemShut {NoStop}%
\bibitem [{\citenamefont {Carter}\ and\ \citenamefont
  {Cross}(2005)}]{carter2005}%
  \BibitemOpen
  \bibfield  {author} {\bibinfo {author} {\bibfnamefont {N.~J.}\ \bibnamefont
  {Carter}}\ and\ \bibinfo {author} {\bibfnamefont {R.~A.}\ \bibnamefont
  {Cross}},\ }\href {\doibase 10.1038/nature03528} {\bibfield  {journal}
  {\bibinfo  {journal} {Nature}\ }\textbf {\bibinfo {volume} {435}},\ \bibinfo
  {pages} {308} (\bibinfo {year} {2005})}\BibitemShut {NoStop}%
\bibitem [{\citenamefont {Schnitzer}\ \emph {et~al.}(2000)\citenamefont
  {Schnitzer}, \citenamefont {Visscher},\ and\ \citenamefont
  {Block}}]{schnitzer2000}%
  \BibitemOpen
  \bibfield  {author} {\bibinfo {author} {\bibfnamefont {M.~J.}\ \bibnamefont
  {Schnitzer}}, \bibinfo {author} {\bibfnamefont {K.}~\bibnamefont {Visscher}},
  \ and\ \bibinfo {author} {\bibfnamefont {S.~M.}\ \bibnamefont {Block}},\
  }\href {\doibase 10.1038/35036345} {\bibfield  {journal} {\bibinfo  {journal}
  {Nature Cell Biology}\ }\textbf {\bibinfo {volume} {2}},\ \bibinfo {pages}
  {718} (\bibinfo {year} {2000})}\BibitemShut {NoStop}%
\bibitem [{\citenamefont {Clemen}\ \emph {et~al.}(2005)\citenamefont {Clemen},
  \citenamefont {Vilfan}, \citenamefont {Jaud}, \citenamefont {Zhang},
  \citenamefont {B\"{a}rmann},\ and\ \citenamefont {Rief}}]{clemen2005}%
  \BibitemOpen
  \bibfield  {author} {\bibinfo {author} {\bibfnamefont {A.~E.}\ \bibnamefont
  {Clemen}}, \bibinfo {author} {\bibfnamefont {M.}~\bibnamefont {Vilfan}},
  \bibinfo {author} {\bibfnamefont {J.}~\bibnamefont {Jaud}}, \bibinfo {author}
  {\bibfnamefont {J.}~\bibnamefont {Zhang}}, \bibinfo {author} {\bibfnamefont
  {M.}~\bibnamefont {B\"{a}rmann}}, \ and\ \bibinfo {author} {\bibfnamefont
  {M.}~\bibnamefont {Rief}},\ }\href {\doibase 10.1529/biophysj.104.053504}
  {\bibfield  {journal} {\bibinfo  {journal} {Biophysical Journal}\ }\textbf
  {\bibinfo {volume} {88}},\ \bibinfo {pages} {4402} (\bibinfo {year}
  {2005})}\BibitemShut {NoStop}%
\bibitem [{\citenamefont {Gennerich}\ \emph {et~al.}(2007)\citenamefont
  {Gennerich}, \citenamefont {Carter}, \citenamefont {{Reck-Peterson}},\ and\
  \citenamefont {Vale}}]{gennerich2007}%
  \BibitemOpen
  \bibfield  {author} {\bibinfo {author} {\bibfnamefont {A.}~\bibnamefont
  {Gennerich}}, \bibinfo {author} {\bibfnamefont {A.~P.}\ \bibnamefont
  {Carter}}, \bibinfo {author} {\bibfnamefont {S.~L.}\ \bibnamefont
  {{Reck-Peterson}}}, \ and\ \bibinfo {author} {\bibfnamefont {R.~D.}\
  \bibnamefont {Vale}},\ }\href {\doibase 10.1016/j.cell.2007.10.016}
  {\bibfield  {journal} {\bibinfo  {journal} {Cell}\ }\textbf {\bibinfo
  {volume} {131}},\ \bibinfo {pages} {952} (\bibinfo {year}
  {2007})}\BibitemShut {NoStop}%
\bibitem [{\citenamefont {Andreasson}\ \emph {et~al.}(2015)\citenamefont
  {Andreasson}, \citenamefont {Milic}, \citenamefont {Chen}, \citenamefont
  {Guydosh}, \citenamefont {Hancock},\ and\ \citenamefont
  {Block}}]{andreasson2015}%
  \BibitemOpen
  \bibfield  {author} {\bibinfo {author} {\bibfnamefont {J.~O.}\ \bibnamefont
  {Andreasson}}, \bibinfo {author} {\bibfnamefont {B.}~\bibnamefont {Milic}},
  \bibinfo {author} {\bibfnamefont {G.}~\bibnamefont {Chen}}, \bibinfo {author}
  {\bibfnamefont {N.~R.}\ \bibnamefont {Guydosh}}, \bibinfo {author}
  {\bibfnamefont {W.~O.}\ \bibnamefont {Hancock}}, \ and\ \bibinfo {author}
  {\bibfnamefont {S.~M.}\ \bibnamefont {Block}},\ }\href {\doibase
  10.7554/eLife.07403} {\bibfield  {journal} {\bibinfo  {journal} {{eLife}}\
  }\textbf {\bibinfo {volume} {4}},\ \bibinfo {pages} {e07403} (\bibinfo {year}
  {2015})}\BibitemShut {NoStop}%
\bibitem [{\citenamefont {Thorn}\ \emph {et~al.}(2000)\citenamefont {Thorn},
  \citenamefont {Ubersax},\ and\ \citenamefont {Vale}}]{thorn2000}%
  \BibitemOpen
  \bibfield  {author} {\bibinfo {author} {\bibfnamefont {K.~S.}\ \bibnamefont
  {Thorn}}, \bibinfo {author} {\bibfnamefont {J.~A.}\ \bibnamefont {Ubersax}},
  \ and\ \bibinfo {author} {\bibfnamefont {R.~D.}\ \bibnamefont {Vale}},\
  }\href {\doibase 10.1083/jcb.151.5.1093} {\bibfield  {journal} {\bibinfo
  {journal} {J Cell Biol}\ }\textbf {\bibinfo {volume} {151}},\ \bibinfo
  {pages} {1093} (\bibinfo {year} {2000})},\ \bibinfo {note} {{PMID:}
  11086010}\BibitemShut {NoStop}%
\bibitem [{\citenamefont {Rai}\ \emph {et~al.}(2013)\citenamefont {Rai},
  \citenamefont {Rai}, \citenamefont {Ramaiya}, \citenamefont {Jha},\ and\
  \citenamefont {Mallik}}]{rai2013}%
  \BibitemOpen
  \bibfield  {author} {\bibinfo {author} {\bibfnamefont {A.~K.}\ \bibnamefont
  {Rai}}, \bibinfo {author} {\bibfnamefont {A.}~\bibnamefont {Rai}}, \bibinfo
  {author} {\bibfnamefont {A.~J.}\ \bibnamefont {Ramaiya}}, \bibinfo {author}
  {\bibfnamefont {R.}~\bibnamefont {Jha}}, \ and\ \bibinfo {author}
  {\bibfnamefont {R.}~\bibnamefont {Mallik}},\ }\href {\doibase
  10.1016/j.cell.2012.11.044} {\bibfield  {journal} {\bibinfo  {journal}
  {Cell}\ }\textbf {\bibinfo {volume} {152}},\ \bibinfo {pages} {172} (\bibinfo
  {year} {2013})}\BibitemShut {NoStop}%
\bibitem [{\citenamefont {Nicholas}\ \emph {et~al.}(2015)\citenamefont
  {Nicholas}, \citenamefont {Berger}, \citenamefont {Rao}, \citenamefont
  {Brenner}, \citenamefont {Cho},\ and\ \citenamefont
  {Gennerich}}]{nicholas2015}%
  \BibitemOpen
  \bibfield  {author} {\bibinfo {author} {\bibfnamefont {M.~P.}\ \bibnamefont
  {Nicholas}}, \bibinfo {author} {\bibfnamefont {F.}~\bibnamefont {Berger}},
  \bibinfo {author} {\bibfnamefont {L.}~\bibnamefont {Rao}}, \bibinfo {author}
  {\bibfnamefont {S.}~\bibnamefont {Brenner}}, \bibinfo {author} {\bibfnamefont
  {C.}~\bibnamefont {Cho}}, \ and\ \bibinfo {author} {\bibfnamefont
  {A.}~\bibnamefont {Gennerich}},\ }\href {\doibase 10.1073/pnas.1417422112}
  {\bibfield  {journal} {\bibinfo  {journal} {Proceedings of the National
  Academy of Sciences}\ }\textbf {\bibinfo {volume} {112}},\ \bibinfo {pages}
  {6371} (\bibinfo {year} {2015})},\ \bibinfo {note} {{PMID:}
  25941405}\BibitemShut {NoStop}%
\bibitem [{\citenamefont {Dudko}\ \emph {et~al.}(2006)\citenamefont {Dudko},
  \citenamefont {Hummer},\ and\ \citenamefont {Szabo}}]{dudko2006}%
  \BibitemOpen
  \bibfield  {author} {\bibinfo {author} {\bibfnamefont {O.~K.}\ \bibnamefont
  {Dudko}}, \bibinfo {author} {\bibfnamefont {G.}~\bibnamefont {Hummer}}, \
  and\ \bibinfo {author} {\bibfnamefont {A.}~\bibnamefont {Szabo}},\ }\href
  {\doibase 10.1103/PhysRevLett.96.108101} {\bibfield  {journal} {\bibinfo
  {journal} {Physical Review Letters}\ }\textbf {\bibinfo {volume} {96}},\
  \bibinfo {pages} {108101} (\bibinfo {year} {2006})}\BibitemShut {NoStop}%
\bibitem [{\citenamefont {Dudko}\ \emph {et~al.}(2008)\citenamefont {Dudko},
  \citenamefont {Hummer},\ and\ \citenamefont {Szabo}}]{dudko2008}%
  \BibitemOpen
  \bibfield  {author} {\bibinfo {author} {\bibfnamefont {O.~K.}\ \bibnamefont
  {Dudko}}, \bibinfo {author} {\bibfnamefont {G.}~\bibnamefont {Hummer}}, \
  and\ \bibinfo {author} {\bibfnamefont {A.}~\bibnamefont {Szabo}},\ }\href
  {\doibase 10.1073/pnas.0806085105} {\bibfield  {journal} {\bibinfo  {journal}
  {Proceedings of the National Academy of Sciences}\ }\textbf {\bibinfo
  {volume} {105}},\ \bibinfo {pages} {15755} (\bibinfo {year} {2008})},\
  \bibinfo {note} {{PMID:} 18852468}\BibitemShut {NoStop}%
\bibitem [{\citenamefont {Klumpp}\ \emph {et~al.}(2015)\citenamefont {Klumpp},
  \citenamefont {Keller}, \citenamefont {Berger},\ and\ \citenamefont
  {Lipowsky}}]{klumpp2015}%
  \BibitemOpen
  \bibfield  {author} {\bibinfo {author} {\bibfnamefont {S.}~\bibnamefont
  {Klumpp}}, \bibinfo {author} {\bibfnamefont {C.}~\bibnamefont {Keller}},
  \bibinfo {author} {\bibfnamefont {F.}~\bibnamefont {Berger}}, \ and\ \bibinfo
  {author} {\bibfnamefont {R.}~\bibnamefont {Lipowsky}},\ }in\ \href
  {http://link.springer.com/chapter/10.1007/978-1-4471-6599-6_3} {\emph
  {\bibinfo {booktitle} {Multiscale Modeling in Biomechanics and
  Mechanobiology}}},\ \bibinfo {editor} {edited by\ \bibinfo {editor}
  {\bibfnamefont {S.}~\bibnamefont {De}}, \bibinfo {editor} {\bibfnamefont
  {W.}~\bibnamefont {Hwang}}, \ and\ \bibinfo {editor} {\bibfnamefont
  {E.}~\bibnamefont {Kuhl}}}\ (\bibinfo  {publisher} {Springer London},\
  \bibinfo {year} {2015})\ pp.\ \bibinfo {pages} {27--61}\BibitemShut {NoStop}%
\bibitem [{\citenamefont {Berger}(2020) }]{berger2020}%
  \BibitemOpen
  \bibfield  {author} {\bibinfo {author} {\bibnamefont {See
        supplementary materials}}\ }\href@noop
  {} {\  \bibinfo {year} {}}\BibitemShut {NoStop}%
\bibitem [{\citenamefont {{DeBerg}}\ \emph {et~al.}(2013)\citenamefont
  {{DeBerg}}, \citenamefont {Blehm}, \citenamefont {Sheung}, \citenamefont
  {Thompson}, \citenamefont {Bookwalter}, \citenamefont {Torabi}, \citenamefont
  {Schroer}, \citenamefont {Berger}, \citenamefont {Lu}, \citenamefont
  {Trybus},\ and\ \citenamefont {Selvin}}]{deberg2013}%
  \BibitemOpen
  \bibfield  {author} {\bibinfo {author} {\bibfnamefont {H.~A.}\ \bibnamefont
  {{DeBerg}}}, \bibinfo {author} {\bibfnamefont {B.~H.}\ \bibnamefont {Blehm}},
  \bibinfo {author} {\bibfnamefont {J.}~\bibnamefont {Sheung}}, \bibinfo
  {author} {\bibfnamefont {A.~R.}\ \bibnamefont {Thompson}}, \bibinfo {author}
  {\bibfnamefont {C.~S.}\ \bibnamefont {Bookwalter}}, \bibinfo {author}
  {\bibfnamefont {S.~F.}\ \bibnamefont {Torabi}}, \bibinfo {author}
  {\bibfnamefont {T.~A.}\ \bibnamefont {Schroer}}, \bibinfo {author}
  {\bibfnamefont {C.~L.}\ \bibnamefont {Berger}}, \bibinfo {author}
  {\bibfnamefont {Y.}~\bibnamefont {Lu}}, \bibinfo {author} {\bibfnamefont
  {K.~M.}\ \bibnamefont {Trybus}}, \ and\ \bibinfo {author} {\bibfnamefont
  {P.~R.}\ \bibnamefont {Selvin}},\ }\href {\doibase 10.1074/jbc.M113.515510}
  {\bibfield  {journal} {\bibinfo  {journal} {The Journal of Biological
  Chemistry}\ }\textbf {\bibinfo {volume} {288}},\ \bibinfo {pages} {32612}
  (\bibinfo {year} {2013})},\ \bibinfo {note} {{PMID:} 24072715 {PMCID:}
  {PMC3820893}}\BibitemShut {NoStop}%
\bibitem [{\citenamefont {Arpa\u{g}}\ \emph {et~al.}(2014)\citenamefont
  {Arpa\u{g}}, \citenamefont {Shastry}, \citenamefont {Hancock},\ and\
  \citenamefont {T\"{u}zel}}]{goekerarpa2014}%
  \BibitemOpen
  \bibfield  {author} {\bibinfo {author} {\bibfnamefont {G.}~\bibnamefont
  {Arpa\u{g}}}, \bibinfo {author} {\bibfnamefont {S.}~\bibnamefont {Shastry}},
  \bibinfo {author} {\bibfnamefont {W.~O.}\ \bibnamefont {Hancock}}, \ and\
  \bibinfo {author} {\bibfnamefont {E.}~\bibnamefont {T\"{u}zel}},\ }\href
  {\doibase 10.1016/j.bpj.2014.09.009} {\bibfield  {journal} {\bibinfo
  {journal} {Biophysical Journal}\ }\textbf {\bibinfo {volume} {107}},\
  \bibinfo {pages} {1896} (\bibinfo {year} {2014})}\BibitemShut {NoStop}%
\bibitem [{\citenamefont {Sumi}(2017)}]{sumi2017}%
  \BibitemOpen
  \bibfield  {author} {\bibinfo {author} {\bibfnamefont {T.}~\bibnamefont
  {Sumi}},\ }\href {\doibase 10.1038/s41598-017-01328-9} {\bibfield  {journal}
  {\bibinfo  {journal} {Scientific Reports}\ }\textbf {\bibinfo {volume} {7}},\
  \bibinfo {pages} {1163} (\bibinfo {year} {2017})}\BibitemShut {NoStop}%
\bibitem [{\citenamefont {Chaudhary}\ \emph {et~al.}(2018)\citenamefont
  {Chaudhary}, \citenamefont {Berger}, \citenamefont {Berger},\ and\
  \citenamefont {Hendricks}}]{chaudhary2018}%
  \BibitemOpen
  \bibfield  {author} {\bibinfo {author} {\bibfnamefont {A.~R.}\ \bibnamefont
  {Chaudhary}}, \bibinfo {author} {\bibfnamefont {F.}~\bibnamefont {Berger}},
  \bibinfo {author} {\bibfnamefont {C.~L.}\ \bibnamefont {Berger}}, \ and\
  \bibinfo {author} {\bibfnamefont {A.~G.}\ \bibnamefont {Hendricks}},\ }\href
  {\doibase 10.1111/tra.12537} {\bibfield  {journal} {\bibinfo  {journal}
  {Traffic}\ }\textbf {\bibinfo {volume} {19}},\ \bibinfo {pages} {111}
  (\bibinfo {year} {2018})}\BibitemShut {NoStop}%
\bibitem [{\citenamefont {Blehm}\ \emph {et~al.}(2013)\citenamefont {Blehm},
  \citenamefont {Schroer}, \citenamefont {Trybus}, \citenamefont {Chemla},\
  and\ \citenamefont {Selvin}}]{blehm2013}%
  \BibitemOpen
  \bibfield  {author} {\bibinfo {author} {\bibfnamefont {B.~H.}\ \bibnamefont
  {Blehm}}, \bibinfo {author} {\bibfnamefont {T.~A.}\ \bibnamefont {Schroer}},
  \bibinfo {author} {\bibfnamefont {K.~M.}\ \bibnamefont {Trybus}}, \bibinfo
  {author} {\bibfnamefont {Y.~R.}\ \bibnamefont {Chemla}}, \ and\ \bibinfo
  {author} {\bibfnamefont {P.~R.}\ \bibnamefont {Selvin}},\ }\href {\doibase
  10.1073/pnas.1219961110} {\bibfield  {journal} {\bibinfo  {journal}
  {Proceedings of the National Academy of Sciences}\ }\textbf {\bibinfo
  {volume} {110}},\ \bibinfo {pages} {3381} (\bibinfo {year} {2013})},\
  \bibinfo {note} {{PMID:} 23404705}\BibitemShut {NoStop}%
\bibitem [{\citenamefont {Berger}\ \emph {et~al.}(2011)\citenamefont {Berger},
  \citenamefont {Keller}, \citenamefont {M\"{u}ller}, \citenamefont {Klumpp},\
  and\ \citenamefont {Lipowsky}}]{berger2011}%
  \BibitemOpen
  \bibfield  {author} {\bibinfo {author} {\bibfnamefont {F.}~\bibnamefont
  {Berger}}, \bibinfo {author} {\bibfnamefont {C.}~\bibnamefont {Keller}},
  \bibinfo {author} {\bibfnamefont {M.~J.}\ \bibnamefont {M\"{u}ller}},
  \bibinfo {author} {\bibfnamefont {S.}~\bibnamefont {Klumpp}}, \ and\ \bibinfo
  {author} {\bibfnamefont {R.}~\bibnamefont {Lipowsky}},\ }\href {\doibase
  10.1042/BST0391211} {\bibfield  {journal} {\bibinfo  {journal} {Biochemical
  Society Transactions}\ }\textbf {\bibinfo {volume} {39}},\ \bibinfo {pages}
  {1211} (\bibinfo {year} {2011})}\BibitemShut {NoStop}%
\bibitem [{\citenamefont {Berger}\ \emph {et~al.}(2012)\citenamefont {Berger},
  \citenamefont {Keller}, \citenamefont {Klumpp},\ and\ \citenamefont
  {Lipowsky}}]{berger2012}%
  \BibitemOpen
  \bibfield  {author} {\bibinfo {author} {\bibfnamefont {F.}~\bibnamefont
  {Berger}}, \bibinfo {author} {\bibfnamefont {C.}~\bibnamefont {Keller}},
  \bibinfo {author} {\bibfnamefont {S.}~\bibnamefont {Klumpp}}, \ and\ \bibinfo
  {author} {\bibfnamefont {R.}~\bibnamefont {Lipowsky}},\ }\href {\doibase
  10.1103/PhysRevLett.108.208101} {\bibfield  {journal} {\bibinfo  {journal}
  {Physical Review Letters}\ }\textbf {\bibinfo {volume} {108}},\ \bibinfo
  {pages} {208101} (\bibinfo {year} {2012})}\BibitemShut {NoStop}%
\bibitem [{\citenamefont {Klumpp}\ and\ \citenamefont
  {Lipowsky}(2005)}]{klumpp2005-a}%
  \BibitemOpen
  \bibfield  {author} {\bibinfo {author} {\bibfnamefont {S.}~\bibnamefont
  {Klumpp}}\ and\ \bibinfo {author} {\bibfnamefont {R.}~\bibnamefont
  {Lipowsky}},\ }\href {\doibase 10.1073/pnas.0507363102} {\bibfield  {journal}
  {\bibinfo  {journal} {Proceedings of the National Academy of Sciences of the
  United States of America}\ }\textbf {\bibinfo {volume} {102}},\ \bibinfo
  {pages} {17284} (\bibinfo {year} {2005})},\ \bibinfo {note} {{PMID:}
  16287974}\BibitemShut {NoStop}%
\bibitem [{\citenamefont {M\"{u}ller}\ \emph {et~al.}(2008)\citenamefont
  {M\"{u}ller}, \citenamefont {Klumpp},\ and\ \citenamefont
  {Lipowsky}}]{mueller2008}%
  \BibitemOpen
  \bibfield  {author} {\bibinfo {author} {\bibfnamefont {M.~J.~I.}\
  \bibnamefont {M\"{u}ller}}, \bibinfo {author} {\bibfnamefont
  {S.}~\bibnamefont {Klumpp}}, \ and\ \bibinfo {author} {\bibfnamefont
  {R.}~\bibnamefont {Lipowsky}},\ }\href {\doibase 10.1073/pnas.0706825105}
  {\bibfield  {journal} {\bibinfo  {journal} {Proceedings of the National
  Academy of Sciences}\ }\textbf {\bibinfo {volume} {105}},\ \bibinfo {pages}
  {4609} (\bibinfo {year} {2008})},\ \bibinfo {note} {{PMID:}
  18347340}\BibitemShut {NoStop}%
\bibitem [{\citenamefont {Berger}\ and\ \citenamefont
  {Hudspeth}(2017)}]{berger2017}%
  \BibitemOpen
  \bibfield  {author} {\bibinfo {author} {\bibfnamefont {F.}~\bibnamefont
  {Berger}}\ and\ \bibinfo {author} {\bibfnamefont {A.~J.}\ \bibnamefont
  {Hudspeth}},\ }\href {\doibase 10.1371/journal.pcbi.1005566} {\bibfield
  {journal} {\bibinfo  {journal} {{PLOS} Computational Biology}\ }\textbf
  {\bibinfo {volume} {13}},\ \bibinfo {pages} {e1005566} (\bibinfo {year}
  {2017})}\BibitemShut {NoStop}%
\bibitem [{\citenamefont {Ruhnow}\ \emph {et~al.}(2017)\citenamefont {Ruhnow},
  \citenamefont {Klo$\beta$},\ and\ \citenamefont {Diez}}]{ruhnow2017}%
  \BibitemOpen
  \bibfield  {author} {\bibinfo {author} {\bibfnamefont {F.}~\bibnamefont
  {Ruhnow}}, \bibinfo {author} {\bibfnamefont {L.}~\bibnamefont {Klo$\beta$}},
  \ and\ \bibinfo {author} {\bibfnamefont {S.}~\bibnamefont {Diez}},\ }\href
  {\doibase 10.1016/j.bpj.2017.09.024} {\bibfield  {journal} {\bibinfo
  {journal} {Biophysical Journal}\ }\textbf {\bibinfo {volume} {113}},\
  \bibinfo {pages} {2433} (\bibinfo {year} {2017})},\ \bibinfo {note} {{PMID:}
  29211997}\BibitemShut {NoStop}%
\end{thebibliography}

\begin{thebibliography}{1}

\bibitem{dembo1994s}
Micah Dembo.
\newblock On peeling an adherent cell from a surface.
\newblock In {\em Lectures on Mathematics in the Life Sciences, Some
  Mathematical Problems in Biology}, pages 51--77. American Mathematical
  Society, Providence, {RI}, 1994.

\bibitem{nicholas2015s}
Matthew~P. Nicholas, Florian Berger, Lu~Rao, Sibylle Brenner, Carol Cho, and
  Arne Gennerich.
\newblock Cytoplasmic dynein regulates its attachment to microtubules via
  nucleotide state-switched mechanosensing at multiple {AAA} domains.
\newblock {\em Proceedings of the National Academy of Sciences},
  112(20):6371--6376, May 2015.
\newblock {PMID:} 25941405.

\bibitem{gillespie1977s}
Daniel~T. Gillespie.
\newblock Exact stochastic simulation of coupled chemical reactions.
\newblock {\em The Journal of Physical Chemistry}, 81(25):2340--2361, December
  1977.

\bibitem{coppin1997s}
Chris~M. Coppin, Daniel~W. Pierce, Long Hsu, and Ronald~D. Vale.
\newblock The load dependence of kinesin's mechanical cycle.
\newblock {\em Proceedings of the National Academy of Sciences of the United
  States of America}, 94(16):8539--8544, August 1997.
\newblock {PMID:} 9238012 {PMCID:} {PMC23000}.

\bibitem{thorn2000s}
Kurt~S. Thorn, Jeffrey~A. Ubersax, and Ronald~D. Vale.
\newblock Engineering the processive run length of the kinesin motor.
\newblock {\em J Cell Biol}, 151(5):1093--1100, November 2000.
\newblock {PMID:} 11086010.

\bibitem{dudko2008s}
Olga~K. Dudko, Gerhard Hummer, and Attila Szabo.
\newblock Theory, analysis, and interpretation of single-molecule force
  spectroscopy experiments.
\newblock {\em Proceedings of the National Academy of Sciences},
  105(41):15755--15760, October 2008.
\newblock {PMID:} 18852468.

\bibitem{deberg2013s}
Hannah~A. {DeBerg}, Benjamin~H. Blehm, Janet Sheung, Andrew~R. Thompson,
  Carol~S. Bookwalter, Seyed~F. Torabi, Trina~A. Schroer, Christopher~L.
  Berger, Yi~Lu, Kathleen~M. Trybus, and Paul~R. Selvin.
\newblock Motor domain phosphorylation modulates kinesin-1 transport.
\newblock {\em The Journal of Biological Chemistry}, 288(45):32612--32621,
  November 2013.
\newblock {PMID:} 24072715 {PMCID:} {PMC3820893}.

\end{thebibliography}
\end{document}